\documentstyle[prd,aps,epsfig,floats]{revtex}
\begin{document}
\draft
\wideabs{
\title{Non-Perturbative U(1) Gauge Theory at Finite Temperature}

\author{Bernd A. Berg$^{\rm \,a,b}$ and Alexei Bazavov$^{\rm \,a,b}$}

\address{ 
$^{\rm \,a)}$ Department of Physics, Florida State University,
  Tallahassee, FL 32306-4350\\
$^{\rm \,b)}$ School of Computational Science, Florida State 
  University, Tallahassee, FL 32306-4120\\
} 

\date{May 29, 2006; revised October 11, 2006} 

\maketitle
\begin{abstract}
For compact U(1) lattice gauge theory (LGT) we have performed a finite 
size scaling analysis on $N_{\tau} N_s^3$ lattices for $N_{\tau}$ fixed 
by extrapolating spatial volumes of size $N_s\le 18$ to $N_s\to\infty$.
Within the numerical accuracy of the thus obtained fits we find for
$N_{\tau}=4$, 5 and~6 second order critical exponents, which exhibit no 
obvious $N_{\tau}$ dependence. The exponents are consistent with 3d 
Gaussian values, but not with either first order transitions or the 
universality class of the 3d XY model. As the 3d Gaussian fixed point 
is known to be unstable, the scenario of a yet unidentified non-trivial 
fixed point close to the 3d Gaussian emerges as one of the possible 
explanations.
\end{abstract}
 }
\narrowtext

\section{Introduction}

Abelian, compact U(1) gauge theory has played a prominent role 
in our understanding of the permanent confinement of quarks. It was  
first investigated by Wilson in his 1974 milestone paper \cite{Wi74}, 
which introduced lattice gauge theory (LGT). For a 4d hypercubic 
lattice his U(1) action reads
\begin{eqnarray} \label{U1action}
  S(\{U\}) = \sum_{\Box} S_{\Box}
\end{eqnarray}
with $S_{\Box}={\rm Re}\left(U_{i_1j_1} U_{j_1i_2} U_{i_2j_2} U_{j_2i_1} 
\right)$, where $i_1,\,j_1,\,i_2$ and $j_2$ label the sites circulating 
about the square $\Box$ and the $U_{ij}$ are complex numbers on the 
unit circle, $U_{ij}=\exp (i\,\phi_{ij})$, $0\le\phi_{ij}<2\pi$.

Wilson concluded that at strong couplings the theory confines static 
test charges due to an area law for the path ordered exponentials 
of the gauge field around closed paths (Wilson loops). A hypothetical 
mechanism of confinement was identified by Polyakov \cite{Po75}, who 
attributed it in 3d Abelian gauge theory to the presence of a monopole 
plasma. For the 4d theory at weak coupling both Wilson and Polyakov 
expected a Coulomb phase in which the test charges are not confined. 
The existence of two distinct phases was later rigorously 
proven~\cite{Gu80}.

So it comes as no surprise that 4d U(1) LGT was the subject of one of 
the very early Monte Carlo (MC) calculations in LGT \cite{CrJa79}. 
One simulates a 4d statistical mechanics with Boltzmann factor
$  \exp \left[ - \beta_g\, S(\{U\}) \right]$
and periodic boundary conditions (other boundary conditions are 
possible too, but are not considered here), $\beta_g=1/g^2$ is 
related to the gauge coupling $g^2$, $\beta_g=0$ is the strong 
and $\beta_g\to\infty$ the weak coupling limit. The study
\cite{CrJa79} allowed to identify the confined and deconfined 
phases. After some debate about the order of the phase transition, 
the bulk transition on symmetric lattices was suggested to be (weakly) 
first order \cite{JeNe83}, a result which was substantiated by 
simulations of the Wuppertal group \cite{ArLi01,ArBu03}. Other 
investigations followed up on the topological properties of the 
theory. This lies outside the scope of the present paper. The 
interested reader may trace this literature from~\cite{Topology}.

The particle excitations of 4d U(1) LGT are called gauge balls and 
in the confined phase also glueballs. Their masses were first studied
in Ref.~\cite{BePa84}. In the confined phase all masses decrease when 
one approaches the transition point. Crossing it, they rise in the 
Coulomb phase with exception of the axial vector mass, which is 
consistent with the presence of a massless photon in that phase. 
Recently this picture was confirmed in Ref.~\cite{MaKo03}, relying on 
far more powerful computers and efficient noise reduction techniques
\cite{LuWe01}. The first order nature of the transition prevents one 
from reaching a continuum limit, as is seen in Fig.~7 of~\cite{MaKo03}. 
In contrast to that investigations in a spherical geometry~\cite{JeLa96} 
and of an extended U(1) Wilson action \cite{CoJe99} reported a scaling 
behavior of glueballs consistent with a second order phase transition. 
But this is challenged in other papers \cite{Isabel,Forcrand}, so that 
it remains questionable whether an underlying non-trivial quantum field 
theory of the confined phase can be defined in this way.

Here we focus on U(1) LGT in finite temperature geometries. We
consider the Wilson action~(\ref{U1action}), choose units $a=1$ 
for the lattice spacing and perform MC simulations on $N_{\tau}N_s^3$ 
lattices. Testing U(1) code for our biased Metropolis-heatbath updating 
(BMHA) \cite{BaBe05}, we noted on small lattices that the characteristics 
of the first order phase transition disappeared when we went from 
the $N_{\tau}=N_s$ to a $N_{\tau}\,N_s^3, \,N_{\tau}<N_s$ geometry. 
This motivated us to embark on a finite size scaling (FSS) calculation 
of the critical exponents of U(1) LGT in the $N_{\tau}\, N_s^3$, 
$N_{\tau}={\rm constant}$, $N_s\to\infty$ geometry. For a review 
of FSS methods and scaling relations see~\cite{PeVi02}.

Later we learned about a paper by Vettorazzo and de Forcrand 
\cite{VeFo04}, who speculate about a scenario of two transitions at
finite, fixed $N_{\tau}$: One for confinement-deconfinement, another 
one into the Coulomb phase, both coinciding only for the zero 
temperature transition. Their claim for the confinement-deconfinement 
transition is that it is first order for $N_{\tau}=8$ and 6, 
$N_s\to\infty$, becoming so weak for $N_{\tau}\le 4$ that it might 
then be second order. In contrast to having two transitions at finite 
$N_{\tau}$ the conventional expectation appears to us one transition, 
which is second order and in the 3d XY universality class, switching
to first order for sufficiently large $N_{\tau}$. See Svetitsky and 
Yaffe \cite{SvYa82} for an early discussion of some of these points.

In the next section we present our numerical results in comparison with
previous literature, followed by summary and conclusions in the final 
section.

\section{Numerical Results}

Our FSS analysis relies on multicanonical simulations \cite{BeNe92} 
for which the parameters were determined using a modification of the 
Wang-Landau (WL) recursion \cite{WaLa01}. A speed up by a factor of 
about three was achieved by implementing the biased Metropolis-Heatbath 
algorithm \cite{BaBe05} for the updating instead of relying on the 
usual Metropolis procedure. This is substantial as, for instance, our 
$16^4$ lattice run takes about 80 days on a 2~GHz PC. Additional 
overrelaxation \cite{Ad81} sweeps were used for some of the simulations.

Our temporal lattice extensions are $N_{\tau}=4,\,5$ and~6. For $N_s$
our values are 4, 5, 6, 8, 10, 12, 14, 16 and~18. Besides we have 
simulated symmetric lattices up to size $16^4$. 
The statistics analyzed in this paper is shown in table~\ref{tab_stat}. 
The lattice sizes are collected in the first and second column. The 
third column contains the number of sweeps spent on the WL recursion 
for the multicanonical parameters. Typically the parameters are frozen 
after reaching $f = e^{1/20}$ for the multiplicative WL factor 
(technical details of our procedure will be published elsewhere). 
Column four lists our production statistics from simulations with 
fixed multicanonical weights. Columns five and six give the $\beta$ 
values between which our Markov process cycled.  Adapting the 
definition of chapter~5.1 of \cite{BBook} one cycle takes the process 
from the configuration space region at $\beta_{\min}$ to $\beta_{\max}$ 
and back. Each run was repeated once more, where after the first run 
the multicanonical parameters were estimated from the statistics of
this run. Columns seven and eight give the number of cycling events 
recorded during runs~1 and~2. 

\begin{table}[tb]
\caption{ Statistics of our MC calculations. The simulation with 
$*$ attached in the WL column uses 22 WL recursions, all others~20.
\label{tab_stat}} 
\centering
\begin{tabular}{|c|c|c|c|c|c|c|c|} 
 & & & & & & \multicolumn{2}{c|}{cycles} \\ \hline 
$L_{\tau}$ & $L$ & WL & sweeps/run  & 
$\beta_{\min}$ & $\beta_{\max}$ & 1 & 2 \\ \hline
 4&  4&18$\,$597& $32\times  20$\,$000$& 0.0  & 1.2  & 213 & 240\\ \hline
 4&  4&11$\,$592& $32\times  20$\,$000$& 0.8  & 1.2  & 527 & 594\\ \hline
 4&  5&14$\,$234& $32\times  12$\,$000$& 0.8  & 1.2  & 146 & 172\\ \hline
 4&  6&19$\,$546& $32\times  32$\,$000$& 0.9  & 1.1  & 258 & 364\\ \hline
 4&  8&29$\,$935& $32\times  32$\,$000$& 0.95 & 1.05 & 229 & 217\\ \hline
 4& 10&25$\,$499& $32\times  64$\,$000$& 0.97 & 1.03 & 175 & 317\\ \hline
 4& 12&47$\,$379&$32\times  112$\,$000$& 0.98 & 1.03 & 338 & 360\\ \hline
 4& 14&44$\,$879&$32\times  112$\,$000$& 0.99 & 1.02 & 329 & 322\\ \hline
 4& 16&54$\,$623&$32\times  128$\,$000$& 0.99 & 1.02 &  19 & 219\\ \hline
 4& 18&58$\,$107&$32\times  150$\,$000$& 0.994& 1.014&  93 & 259\\ \hline
 5&  5&18$\,$201& $32\times  12$\,$000$& 0.8  & 1.2  & 114 & 122\\ \hline
 5&  6&20$\,$111& $32\times  36$\,$000$& 0.9  & 1.1  & 294 & 308\\ \hline
 5&  8&31$\,$380& $32\times  40$\,$000$& 0.95 & 1.05 &  35 & 191\\ \hline
 5& 10&47$\,$745& $32\times  72$\,$000$& 0.97 & 1.03 & 144 & 231\\ \hline
 5& 12&37$\,$035& $32\times 112$\,$000$& 0.99 & 1.02 & 280 & 326\\ \hline
 5& 14&49$\,$039& $32\times 112$\,$000$& 1.0  & 1.02 & 192 & 277\\ \hline
 5& 16&43$\,$671& $32\times 160$\,$000$& 1.0  & 1.02 & 226 & 257\\ \hline
 5& 18&56$\,$982&$32\times  180$\,$000$& 1.0  & 1.014& 138 & 241\\ \hline
 6&  6&28$\,$490& $32\times  40$\,$000$& 0.9  & 1.1  & 312 & 281\\ \hline
 6&  8&44$\,$024& $32\times  40$\,$000$& 0.96 & 1.04 & 173 & 175\\ \hline
 6& 10&51$\,$391& $32\times  72$\,$000$& 0.97 & 1.04 & 139 & 170\\ \hline
 6& 12&41$\,$179& $32\times 128$\,$000$& 0.995& 1.02 & 226 & 283\\ \hline
 6& 14&50$\,$670& $32\times 128$\,$000$& 1.0  & 1.02 &  89 & 220\\ \hline
 6& 16&56$\,$287& $32\times 160$\,$000$& 1.0  & 1.02 & 149 & 189\\ \hline
 6& 18&68$\,$610& $32\times  180$\,$000$&1.005& 1.015& 123 & 200\\ \hline
 8&  8&46$\,$094& $32\times  40$\,$000$& 0.97 & 1.03 & 111 & 159\\ \hline
10& 10&48$\,$419& $32\times  96$\,$000$& 0.98 & 1.03 & 103 & 133\\ \hline
12& 12&70$\,$340& $32\times 112$\,$000$& 0.99 & 1.03 &  75 &  82\\ \hline
14&14&112$\,$897& $32\times 128$\,$000$& 1.0  & 1.02 &  57 &  51\\ \hline
16& 16&87$\,$219& $32\times 160$\,$000$& 1.007& 1.015&  12 &  73\\ \hline
16&16&191$\,$635*&$32\times 160$\,$000$& 1.007& 1.015&  48 &  74\\
\end{tabular} \end{table} 

Using the logarithmic coding of chapter~5.1.5 of \cite{BBook} physical 
observables are reweighted to canonical ensembles. Error bars as shown 
in figures are calculated using jackknife bins (e.g., chapter~2.7 of 
\cite{BBook}) with their number given by the first value in column 
four (always 32), while the second value was also used for the number 
of equilibrium sweeps (without measurements) performed after the 
recursion. Weighted by the number of their completed cycles, the 
results from two or more runs are combined for for the final 
analysis (compare chapter~2.1.2 of \cite{BBook}).

\subsection{Action variables}

\begin{figure}[-t] \begin{center} 
\epsfig{figure=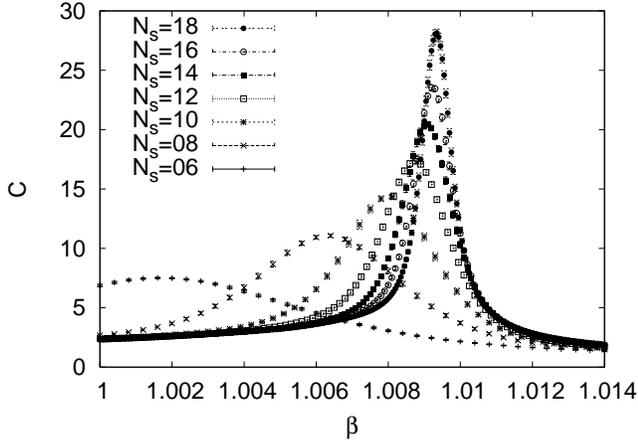,width=\columnwidth} \vspace{-1mm}
\caption{Finite size dependence of the specific heat functions 
$C(\beta)$ on $N_{\tau}=6$ lattices. \label{fig_C06} }
\end{center} \vspace{-3mm} \end{figure}

\begin{figure}[-t] \begin{center} 
\epsfig{figure=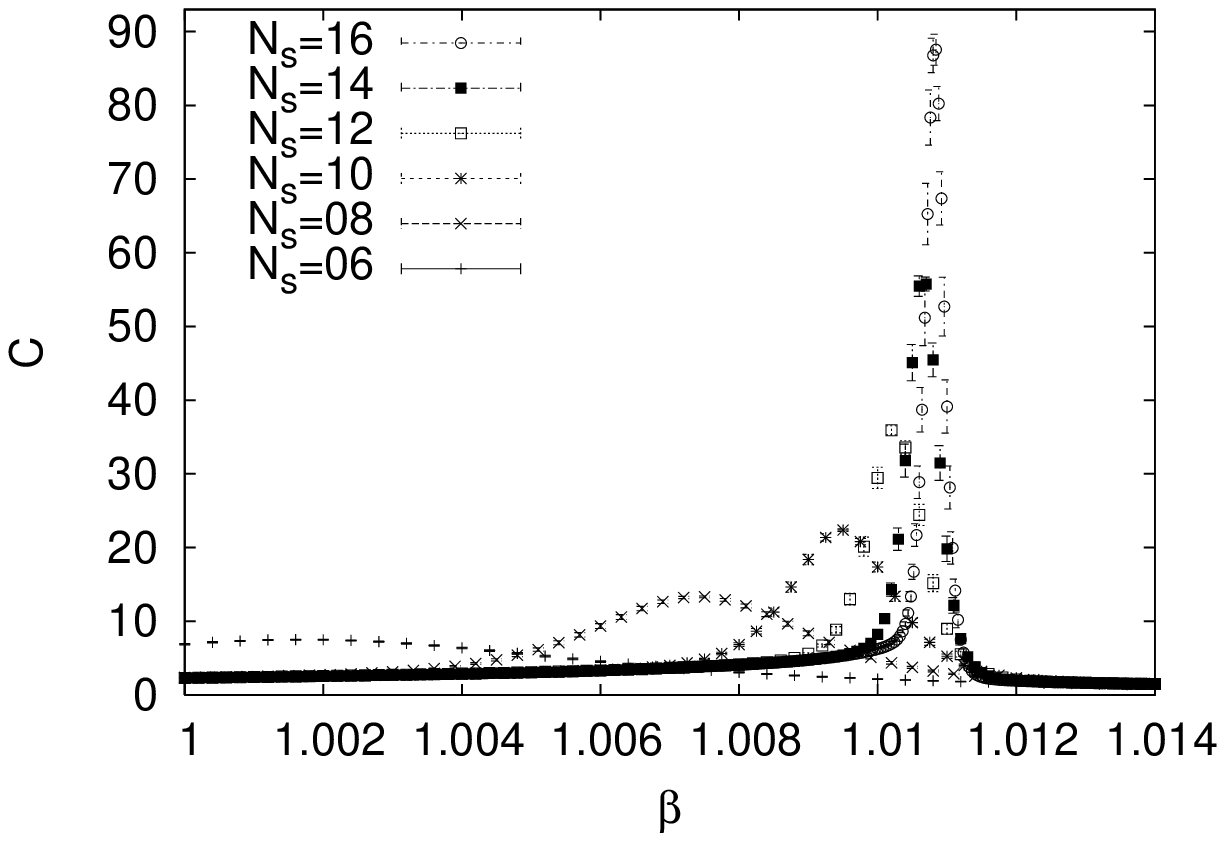,width=\columnwidth} \vspace{-1mm}
\caption{Finite size dependence of the specific heat functions 
$C(\beta)$ on $N_{\tau}=N_s$ lattices. 
\label{fig_C00} }
\end{center} \vspace{-3mm} \end{figure}

Figures~\ref{fig_C06} and~\ref{fig_C00} show for various values of 
$N_s$ the specific heat 
\begin{equation} \label{Cbeta}
  C(\beta)\ =\ \frac{1}{6N} \left[\langle S^2\rangle -
  \langle S\rangle^2 \right]~~{\rm with}~~ N = N_{\tau}\,N_s^3
\end{equation} 
in the neighborhood of the phase transition for $N_{\tau}=6$ and on 
symmetric lattices. The $\beta$ ranges in the figures are chosen to 
match. 

\begin{figure}[-t] \begin{center} 
\epsfig{figure=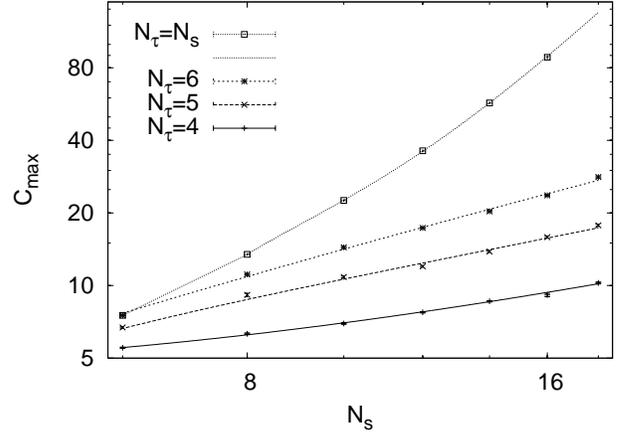,width=\columnwidth} \vspace{-1mm}
\caption{Maxima of the specific heat. \label{fig_Cmax} }
\end{center} \vspace{-3mm} \end{figure}

In  Fig.~\ref{fig_Cmax} we show all our specific heat maxima on a 
log-log scale. Our data for the symmetric lattices are for $N_s\ge 8$ 
consistently described by a fit to the first order transition form 
\cite{BK92} $C_{\max}(N_s)/(6N)=c_0+a_1/N+a_2/N^2$. The goodness of our 
fit is $Q=0.64$ (see, e.g., chapter~2.8 of Ref.~\cite{BBook} for the 
definition and a discussion of $Q$), and its estimate for the specific 
heat density is $c_0=0.0001961\,(26)$. This is 10\% higher than the 
$c_0$ value reported by the Wuppertal group \cite{ArBu03}, where 
lattices up to size $18^3$ were used. For the interface tension 
consistent fits to a small positive as well as to a zero infinite 
volume value are possible. Interestingly the Wuppertal group decided 
against including their largest lattices from \cite{ArLi01} into their 
final analysis \cite{ArBu03}, because action data on them did not cover 
the double peak region coherently (communicated by T. Neuhaus).

For $N_{\tau} = 4,\,5$ and~6 our curves in Fig.~\ref{fig_Cmax} are 
linear fits in $N_s$, $C_{\max}(N_s)=a_1N_s+a_0+a_{-1}/N_s.$ For 
$N_{\tau}=4$ the goodness of this fit is $Q=0.20$ using our 
$N_s\ge 6$ data. But for $N_{\tau}=5$ and~6 the $Q$ values are 
unacceptably small, although the data scatter nicely about the 
curves. For large $N_s$ the maxima of the specific heat curves scale 
like (see \cite{PeVi02})
\begin{equation} \label{Cmax}
  C_{\max}(N_s)\ \sim\ N_s^{\alpha/\nu}\,,
\end{equation} 
where one has $\alpha/\nu=4$ in case of the first order transition
for $N_{\tau}=N_s$. In the $N_{\tau}$ fixed, $N_s\to\infty$ geometry 
the systems become three-dimensional, so that $\alpha/\nu =3$ would 
be indicative of a first order transition, while our data are 
consistent with the second order exponent $\alpha/\nu =1$.

This has to be contrasted with the claim by Vettorazzo and de Forcrand 
\cite{VeFo04} that the $N_{\tau}\ge 6$ transitions are first order.
For $N_{\tau}=8$ and~6 their evidence relies on simulations of very 
large lattices. Differences in action values obtained after ordered 
and disordered starts support a non-zero latent heat in the infinite 
volume limit. For $N_{\tau}=6$ the spatial lattice sizes used are 
$N_s=48$ and~60 and their MC statistics shown consists of 5$\,$000 
measurements per run, separated by one heatbath plus four overrelaxation 
sweeps (these units are not defined in \cite{VeFo04}, but were 
communicated to us by de Forcrand and previously used in 
\cite{Forcrand}). For a second order phase transition the integrated 
autocorrelation time $\tau_{\rm int}$ scales approximately $\sim N_s^2$ 
and we estimate from our own simulations on smaller lattices that in 
units of those measurements $\tau_{\rm int} \approx 7$\,$000$ for 
$N_{\tau}=6$ and $N_s=48$. A MC segment of the length of 
$\tau_{\rm int}$ delivers one statistically independent event (e.g.,
chapter~4.1.1 of \cite{BBook}). Therefore, the run of \cite{VeFo04} 
would in case of a second order transition be based on less than one 
event and strong metastabilities would be expected as soon as the 
Markov chain approaches the scaling region. For $N_s=60$ and the 
$N_{\tau}=8$ lattices the situation is even worse. We conclude that 
these data cannot decide the order of the transition. 

Let us remind the reader that a double peak alone does not signal a 
first order transition. One has to study its FSS behavior, but no error 
bars can be estimated when one has only one statistically independent 
event.  Actually for our larger spatial volumes we find double peaks 
in our $6\times N_s^3$ action histograms and they are also well-known 
to occur for the magnetization of the 3d Ising model at its critical 
point \cite{Bi81}. 

\subsection{Polyakov loop variables}

Besides the action we measured Polyakov loops and their low-momentum 
structure factors. For U(1) LGT Polyakov loops are the $U_{ij}$ 
products along the straight lines in $N_{\tau}$ direction. Each 
Polyakov loop $P_{\vec{x}}$ is a complex number on the unit circle, 
which depends only on the space coordinates, quite like a XY spin in 
3d. We calculate the sum over all Polyakov loops on the lattice
\begin{equation} \label{PLoop}
  P\ =\ \sum_{\vec{x}} P_{\vec{x}}\,.
\end{equation} 
The critical exponent $\gamma/\nu$ is obtained from the maxima of the 
susceptibility of the absolute value $|P|$,
\begin{equation} \label{chi}
  \chi_{\max}\ =\ \frac{1}{N_s^3} \left[ \langle|P|^2\rangle -
  \langle|P|\rangle^2\ \right]_{\max} \sim\ N_s^{\gamma/\nu}\,,
\end{equation} 
and $(1-\beta)/\nu$ from the maxima of
\begin{equation} \label{chi_beta}
  \chi^{\beta}_{\max}\ =\ \frac{1}{N_s^3} \left. \frac{d~}{d\beta}\,
  \langle|P|\rangle\ \right|_{\max} \sim\ N_s^{(1-\beta)/\nu}\,.
\end{equation} 
Structure factors are defined by (see, e.g., Ref.~\cite{Stanley})
\begin{equation} \label{sf}
  F(\vec{k})=\frac{1}{N_s^3} \left\langle\left|\sum_{\vec{r}}
  P(\vec{r})\, \exp(i\vec{k}\vec{r}) \right|^2\right\rangle\,,
  \vec{k} = \frac{2\pi}{N_s}\vec{n}\,,
\end{equation} 
where $\vec{n}$ is an integer vector, which is for our measurements 
restricted to $(0,0,1)$, $(0,1,0)$, and $(1,0,0)$. Maxima of structure 
factors scale like
\begin{equation} \label{sffss}
  F_{\max}(\vec{k})\ \sim\ N_s^{2-\eta}\,.
\end{equation} 

The exponents can be estimated from two parameter fits (A)
$Y=a_1\,N_s^{a_2}.$ Due to finite size corrections the goodness $Q$ 
of these fits will be too small when all lattice sizes are included. 
The strategy is then not to overweight~\cite{overweight} the small 
lattices and to omit, starting with the smallest, lattices altogether 
until an acceptable $Q\ge 0.05$ has been reached. We found a rather 
slow convergence of the thus obtained estimates with increasing lattice 
size. This can improve by including more parameters in the fit. So we 
used the described strategy also for three parameter fits (B) 
$Y=a_0+a_1\,N_s^{a_2}.$ The penalty for including more parameters is 
in general increased instability against fluctuations of the data and, 
in particular, their error bars. For a number of our data sets this 
is the case for fit~B, so that an extension to more than three 
parameters makes no sense. We performed first the fit~B for each 
data set, but did fall back to fit~A when no consistency or stability 
was reached for a fit~B including at least the five largest lattices. 
The thus obtained values are listed in table~\ref{tab_exponents}. 
Table~\ref{tab_fitinfo} gives additional information about the fits.

\begin{table}[tb]
\caption{ Estimates of critical exponents as explained in the text.
Properties of the fits are summarized in table~\ref{tab_fitinfo}.
\label{tab_exponents}} 
\centering
\begin{tabular}{|c|c|c|c|c|} 
$N_{\tau}$ & $\alpha/\nu$ & $\gamma/\nu$ &
             $(1-\beta)/\nu$  & $2-\eta$\\ \hline
 4 & 1.15 (10) & 1.918 (34) & 1.39 (7) & 1.945 (10) \\ \hline
 5 & 0.97 (04) & 2.086 (79) & 1.51 (4) & 1.955 (20) \\ \hline
 6 & 1.31 (07) & 1.968 (37) & 1.59 (4) & 1.901 (31) \\ \hline
n-t&1.15 (15)& 1.95  (5) & 1.55 (5) & 1.95 (5) \\ 
\end{tabular} \end{table} 

\begin{table}[tb]
\caption{ Number of data used and type of fit (A or B as 
explained in text), goodness of fit $Q$.  \label{tab_fitinfo}} 
\centering
\begin{tabular}{|c|c|c|c|c|} 
$N_{\tau}$ & $\alpha/\nu$ & $\gamma/\nu$ &
             $(1-\beta)/\nu$  & $2-\eta$\\ \hline
 4 & 7B, 0.25 & 7B, 0.21 & 7B, 0.25  & 8B, 0.78 \\ \hline
 5 & 4A, 0.76 & 6B, 0.40 & 4A, 0.80  & 7B, 0.23 \\ \hline
 6 & 3A, 0.09 & 7B, 0.09 & 4A, 0.83  & 5B, 0.42 \\ 
\end{tabular} \end{table} 

\begin{figure}[-t] \begin{center} 
\epsfig{figure=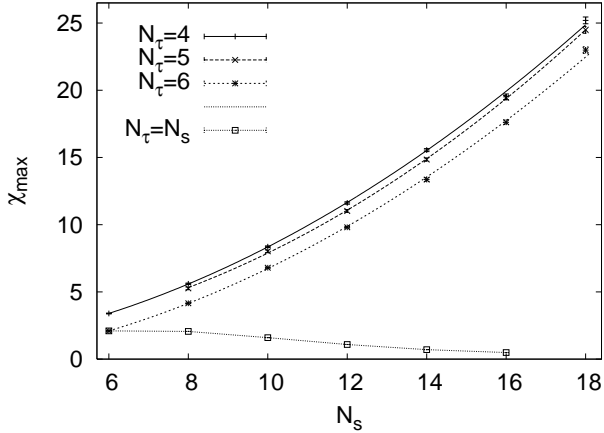,width=\columnwidth} \vspace{-1mm}
\caption{Maxima of Polyakov loop susceptibilities.
\label{fig_chi} }
\end{center} \vspace{-3mm} \end{figure}

Our lattices support second order transitions for $N_{\tau}=4,\,5$ 
and~6. The evidence is best for observables derived from Polyakov 
loops. For example, in Fig.~\ref{fig_chi} we show our data for the 
maxima of the Polyakov loops susceptibility together with their fits 
used in table~\ref{tab_exponents} (for the 
symmetric lattices the data are connected by straight lines).
For fixed $N_{\tau}$ we find an approximately quadratic increase 
with $N_s$, while there is a decrease for the symmetric lattices, 
which appears to converge towards zero or a finite discontinuity
(note that one has no common scale for Polyakov loops from symmetric
lattices, because their lengths change with $N_{\tau}$).

\begin{figure}[-t] \begin{center} 
\epsfig{figure=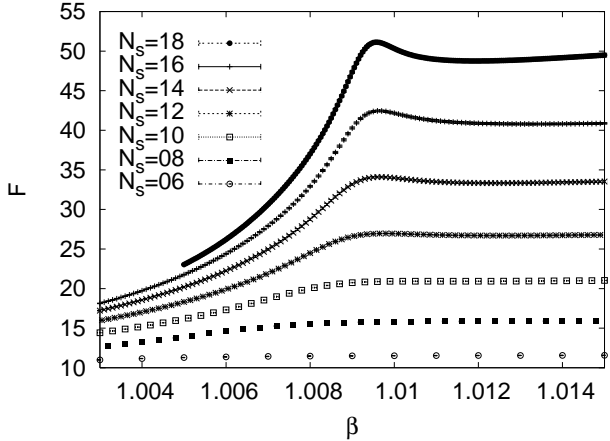,width=\columnwidth} \vspace{-1mm}
\caption{Structure factors (\ref{sf}) for $N_{\tau}=6$.
\label{fig_sf06} }
\end{center} \vspace{-3mm} \end{figure}

Our structure factor data support that one is for $\beta>\beta_c$ in
the Coulomb phase: As shown for $N_{\tau}=6$ in Fig.~\ref{fig_sf06}
the structure factors remain divergent for $\beta>\beta_c$, as 
expected for a power law fall-off of Polyakov loop correlations. These 
observations apply to the $\beta$ ranges (compare table~\ref{tab_stat}) 
covered by our multicanonical simulations. To have still reasonably 
many cycling events on large lattices, this range was chosen to shrink
with increasing lattice size. So we test not very far into the $\beta
> \beta_c$ phase.

The Polyakov loops describe 3d spin systems. So one would like to 
identify whether the observed transitions are in any of their known 
universality classes. At first thought the universality class of the 
3d XY model comes to mind (e.g., \cite{SvYa82}), because the symmetry 
is correct. It is easy to see that the $N_{\tau}=1$ gauge system 
decouples into a 3d XY model and a 3d U(1) gauge theory. The latter 
has no transition and is always confined. But one cannot learn
much from this observation as there is no interaction between
the two systems.
Surprisingly the data of table~\ref{tab_exponents} do not support 
the XY universality class. Although our estimates of $\gamma/\nu$ 
agree with what is expected, $\alpha/\nu$ is entirely off. For the 
XY model a small negative value is established~\cite{PeVi02}, while 
Fig.~\ref{fig_Cmax} shows that all our specific heat maxima increase 
steadily. We remark that the scenario may change for $N_{\tau}<4$. 
We have preliminary results for $N_{\tau}=2$ and~3.
The increase of the specific heat maxima becomes considerably weaker 
than for $N_{\tau}=4$. For $N_{\tau}=2$ it slows continuously down
with increasing lattice size (so far up to $N_s=20$) and one can 
imagine that it comes altogether to a halt. Once completed, our 
simulations for $N_{\tau}=2$ and~3 will be reported elsewhere. 

\section{Summary and Conclusions}

In view of expected systematic errors due to our limited lattice sizes, 
one can state that our estimates of table~\ref{tab_exponents} are 
consistent with the Gaussian values $\alpha/\nu=1$ and $\gamma/\nu=2$
(with error bars 0.3 for $\alpha/\nu$ and 0.1 for $\gamma/\nu$).
Using the hyperscaling relation $2-\alpha=d\nu$ with $d=3$ yields 
$\alpha=\nu=1/2$. The other estimates of exponents listed in 
table~\ref{tab_exponents} provide consistency checks as they 
are linked to $\alpha/\nu=1$ and $\gamma/\nu=2$ by the scaling 
relations $\alpha+2\beta+\gamma=2$ and $\gamma/\nu=2-\eta$. For
the Gaussian exponents $(1-\beta)/\nu=1.5$ and $\eta=0$ follows, 
both consistent with the data of the table.

However, the problem with the Gaussian scenario is that the Gaussian 
renormalization group fixed point in 3d has two relevant operators
\cite{Gaussian}. So one does not understand why the effective spin 
system should care to converge into this fixed point \cite{SvYa82}. 
Therefore, the interesting scenario of a new non-trivial (n-t) fixed 
point with exponents accidentally close to 3d Gaussian arises. An 
illustration, which is consistent with the data, is given in the last 
row of table~\ref{tab_exponents}. The mean values are constructed to 
fulfill the scaling relations and match with $\nu=0.482$, 
$\alpha=0.554$, $\gamma=0.94$, $\beta=0.253$, $\eta=0.05$. 

\begin{figure}[-t] \begin{center} 
\epsfig{figure=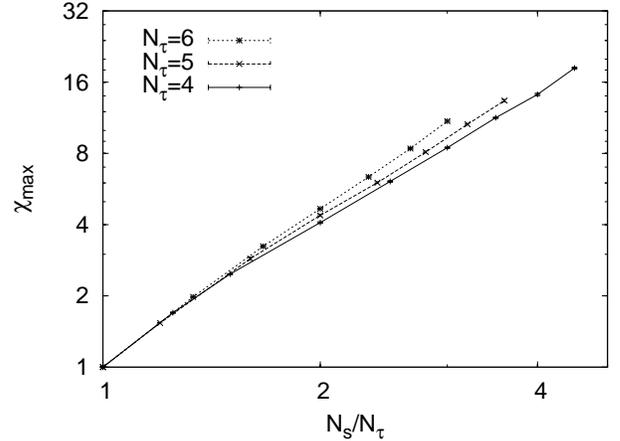,width=\columnwidth} \vspace{-1mm}
\caption{Rescaled maxima of Polyakov loops susceptibilities. 
\label{fig_CPnorm} }
\end{center} \vspace{-3mm} \end{figure}

One may expect that the first order transition of the symmetric 
lattices prevails once $N_{\tau}$ is larger than the correlation 
length on symmetric lattices. But a non-zero interface tension has 
never been established for this transition. So one could also
imagine an instability under the change of the geometry. From a FSS 
point of view it appears then natural that the character of the 
transition will not change anymore, once a value of $N_{\tau}$ has 
been reached, which is sufficiently large to be insensitive to 
lattice artifacts. Up to normalizations data from $N_{\tau}N_s^3$ 
and $2N_{\tau}(2N_s)^3$, $N_s>N_{\tau}$ lattices should then become 
quite similar. We illustrate this here by rescaling the maxima of 
our Polyakov loop susceptibilities with a common factor, so that 
they become equal to~1 on symmetric lattices. On a log-log scale 
the results are then plotted in Fig.~\ref{fig_CPnorm} against 
$N_s/N_{\tau}$. The behavior is consistent with assuming a common 
critical exponent for all of them (parallel lines are then expected 
for large $N_s/N_{\tau}$). 

The litmus test for identifying a second order phase transition
is that one is able to calculate its critical exponents unambiguously.
Instead of starting with data of uncontrolled quality from very large 
lattices, the FSS strategy is to control finite size effects by working 
the way up from small to large systems. With MC calculations FSS methods 
find their limitations through the lattice sizes, which fit into the 
computer and can be accurately simulated in a reasonable time. Within 
the multicanonical approach ``accurately'' means that one has to get 
the system cycling through the entire critical or first order region,
and at least about one hundred cycles ought to be completed with
measurements.

Our lattice sizes are not small on 
the scale of typical numerical work on U(1) LGT, for instance the 
lattices used for the Wuppertal $c_0$ estimate of \cite{ArBu03}. 
But we have not yet reached lattices large enough to provide hard 
evidence that there is no $N_s\to\infty$ turn-around towards either 
a first order transition or the 3d XY fixed point. In particular
in view of the fact that our data do not support the generally 
expected scenario, it would be desirable to extent the present 
analysis to the largest lattices that can be reached by extensive 
simulations on supercomputers, instead of relying on relatively 
small PC clusters.

With mass spectrum methods \cite{MaKo03,Ta06} one may investigate 
the scaling behavior of the model from a different angle. In particular 
observation of a massless photon \cite{BePa84} can provide more direct
evidence for the Coulomb phase than our structure factor measurements.
Finally, renormalization group theory could contribute to clarifying 
the issues raised by our data. Amazingly, even after more than thirty 
years since Wilson's paper \cite{Wi74} the nature of U(1) LGT phase
transition is still not entirely understood.

\acknowledgments
We thank Urs Heller for checking on one of our action distributions 
with his own code and Thomas Neuhaus for communicating details of
the Wuppertal data. We are indebted to Philippe de Forcrand for e-mail 
exchanges and useful discussions. This work was in part supported 
by the US Department of Energy under contract DE-FA02-97ER41022. 
Our data were generated on PC clusters at FSU.

\clearpage

\begin{thebibliography}{19}

\bibitem{Wi74} K.G. Wilson, 
               Phys. Rev. D {\bf 10}, 2445 (1974). 

\bibitem{Po75} A.M. Polyakov, 
               Phys. Lett. B {\bf 59}, 82 (1975). 

\bibitem{Gu80} A.H. Guth, 
              Phys. Rev. D {\bf 21}, 2291 (1980).

\bibitem{CrJa79} M. Creutz, L. Jacobs, and C. Rebbi, Phys. Rev. D
                 {\bf 20}, 1915 (1979). 

\bibitem{JeNe83} J. Jersak, T. Neuhaus, and P.M. Zerwas, Phys. Lett. 
                 B {\bf 133}, 103 (1983). 

\bibitem{ArLi01} G. Arnold, T. Lippert, T. Neuhaus, and K. Schilling, 
                 Nucl. Phys. B (Proc. Suppl.) {\bf 94}, 651 (2001) 
                 [hep-lat/0011058].

\bibitem{ArBu03} G. Arnold, B. Bunk, T. Lippert, and K. Schilling,
                 Nucl. Phys. B (Proc. Suppl.) {\bf 119}, 864 (2003)
                 [hep-lat/0210010] and references given therein.

\bibitem{Topology} Y. Koma, M. Koma, and P. Majumdar, Nucl. Phys. B
                   {\bf 692}, 209 (2004); M. Panero, JHEP 05, 66 
                   (2005).

\bibitem{BePa84} B.A. Berg and C. Panagiotakopoulos, Phys. Rev. Lett.
                 {\bf 52}, 94 (1984).

\bibitem{MaKo03} P. Majumdar, Y. Koma, and M. Koma, 
                 Nucl. Phys. B {\bf 677}, 273 (2004) [hep-lat/0309003]. 

\bibitem{LuWe01} L. L\"uscher and P. Weisz, JHEP 09, 10 (2001)
                 [hep-lat/0207003].

\bibitem{JeLa96} J. Jersak, C.B. Lang, and T. Neuhaus, Phys. Rev. D
                 {\bf 54}, 6909 (1996) [hep-lat/9606013] 

\bibitem{CoJe99} J. Cox, W. Franzki, J. Jersak, H. Pfeiffer, C.B. Lang,
                 T. Neuhaus, and P.W.  Stephenson, Nucl. Phys.  B {\bf 
                 499}, 607 (1997) [hep-lat/9701005];
                 J. Cox, J. Jersak, H. Pfeiffer, T. Neuhaus, P.W. 
                 Stephenson, and A. Seyfried, 
                 Nucl. Phys. B {\bf 545}, 607 (1999) [hep-lat/9808049]
                 and references given therein.

\bibitem{Isabel} I. Campos, A. Cruz, and A. Taranc\'on, Nucl. Phys.
                 B {\bf 528}, 325 (1998) [hep-lat/9803007].  

\bibitem{Forcrand} M. Vettorazzo and P. de Forcrand, Nucl. Phys.
                   B {\bf 668}, 85 (2004) [hep-lat/0311006]. 

\bibitem{BaBe05} A. Bazavov and B.A. Berg, Phys. Rev. D {\bf 71},
                 114506 (2005)

\bibitem{PeVi02} A. Pelissetto and E. Vicari, 
                 Phys. Rep. {\bf 368}, 549 (2002) [cond-mat/0012164].

\bibitem{VeFo04} M. Vettorazzo and P. de Forcrand, Phys. Lett. B
                {\bf 604}, 82 (2004) [hep-lat/0409135]. 

\bibitem{BeNe92} B.A. Berg, and T. Neuhaus, Phys. Rev. Lett. {\bf 68}, 
                 9 (1992).

\bibitem{WaLa01} F. Wang and D.P. Landau, Phys. Rev. Lett {\bf 86}, 
                 2050 (2001).

\bibitem{SvYa82} B. Svetitsky and L.G. Yaffe, Nucl. Phys. B {\bf 210},
                 423 (1982).

\bibitem{Ad81} S.L. Adler, Phys. Rev. D {\bf 37}, 458 (1981).

\bibitem{BBook} B.A. Berg, {\it Markov Chain Monte Carlo Simulations
and Their Statistical Analysis}, World Scientific, Singapore, 2004.

\bibitem{BK92} C. Borgs and R. Kotecky, Phys. Rev. Lett {\bf 68}, 1734
               (1992) and references given therein.

\bibitem{Bi81} K. Binder, Z. Phys. B: Condens. Matter 
               {\bf 43}, 1699 (1981).

\bibitem{Stanley} H.E. Stanley, {\it Introduction to Phase Transitions
and Critical Phenomena}, Clarendon Press, Oxford, 1971, pp.~98.

\bibitem{overweight} As MC runs on small lattices are ``cheap'', one
tends to have for them more accurate data than for the valuable large 
lattices. Care should be taken that the error to signal ratios are
similar for all included lattice sizes.

\bibitem{Gaussian} E.g., F.J. Wegner, in {\it Phase Transitions and
                   Critical Phenomena}, C. Domb and M.S. Green editors
                   (Academic Press, New York, 1976), p.59.

\bibitem{Ta06} L.~Tagliacozzo, hep-lat/0603022, Phys. Lett. B, to 
               appear.

\end{thebibliography}
\end{document}